\documentclass[a4paper,fleqn]{an}
\usepackage{amsmath}
\usepackage{graphicx}
\usepackage{times}
\usepackage{fancyhdr}
\usepackage{natbib}

\def\citeq#1{(\ref{#1})}
\def\d{\text{d}}
\def\D#1{\nabla_{\!#1}}

\def\-{\ifmmode \!-\! \else \mbox{\bf --}\fi}
\def\={\ifmmode \!=\! \else \mbox{=}\fi}
\def\ee#1{\ifmmode {}\!\times\!10^{#1} \else ${}\!\times\!10^{#1}$\fi}
\def\about{\ifmmode \sim\, \else $\sim$\,\fi}
\def\ratio#1#2{{{^{#1}}\!/{_{#2}}}}
\def\deg{^\circ}

\def\kB{k_\text{B}}
\def\amu{m_\text{u}}
\def\mw{\mu_{\text w}}
\def\OmegaK{\Omega_\text{K}}
\def\ellK{\ell_\text{K}}
\def\TK{T_\text{K}}

\def\Ueff{\mathcal{U}}
\def\inc{i}

\def\Hz{\,\text{Hz}}

\def\cm{\,\text{cm}}

\def\s{\,\text{s}}

\def\erg{\,\text{erg}}
\def\finestruct{\alpha_{\scriptscriptstyle f}}
\def\sigmathomson{\sigma_{\scriptscriptstyle T}}

\def\ie{{\it i.e.\ }}
\def\eg{{\it e.g.\ }}
\def\cf{{\it c.f.\ }}

\begin{document}

%
%
%
%
%

\Issue{9}		
\Volume{326}            
\Yearsubmission{2005}   
\Yearpublication{2005}  
\Pagespan{849}{855}         
\DOI{10426}		

\headnote{Astron. Nachr. {\bf /} AN {\bf \volume}, No. \issue, \pages\
(\yearofpublic) {\bf / DOI} 10.1002/asna.\yearofsubmission\doi}

%
\fancyhead{}
\fancyhead[LE,RO]{\small\thepage}
\fancyhead[RE]{\small Astron. Nachr. {\bf /} AN~{\bf \volume}, 
No. \issue\ (\yearofpublic) {\bf /} www.an-journal.org}
\fancyhead[LO]{M. Bursa: 
Global oscillations of a fluid torus as a modulation mechanism
for black-hole high-frequency QPOs}

%
\fancyfoot{}
\fancyfoot[LE,RO]{%
\scriptsize%
\copyright\/\yearofpublic\ WILEY-VCH Verlag GmbH \& Co. KGaA, Weinheim}

\title{Global oscillations of a fluid torus as a modulation mechanism for black-hole high-frequency QPOs}

\author{M. Bursa}
\institute{Astronomical Institute, Academy of Sciences, Bo\v{c}n\'{\i} II 1401, 141\,31~Praha 4, Czechia}

\date{Received 8 September 2005; accepted 22 September 2005; 
published online 20 October 2005}

\abstract{
We study strong-gravity effects on modulation of radiation emerging from accreting compact objects as a possible mechanism for flux modulation in QPOs. We construct a toy model of an oscillating torus in the slender approximation assuming thermal bremsstrahlung for the intrinsic emissivity of the medium and we compute observed (predicted) radiation signal including contribution of indirect (higher-order) images and caustics in the Schwarzschild spacetime. We show that the simplest oscillation mode in an accretion flow, axisymmetric up-and-down motion at the meridional epicyclic frequency, may be directly observable when it occurs in the inner parts of accretion flow around black holes. Together with the second oscillation mode, an in-and-out motion at the radial epicyclic frequency, it may then be responsible for the high-frequency modulations of the \mbox{X-ray} flux observed at two distinct frequencies (twin HF-QPOs) in micro-quasars.
}

\correspondence{bursa@astro.cas.cz}

\maketitle

\section{Introduction}

X-ray radiation coming from accreting black hole binary sources can show quasi-periodic modulations at two distinct high frequencies \mbox{($>30\Hz$)}, which appear in the \mbox{$3:2$} ratio \citep{McClintockRemillard05}. Observations show that the solely presence of a thin accretion disk is not sufficient to produce these HFQPO modulations, because they are exclusively connected to the spectral state, where the energy spectrum  is dominated by a steep power law with some weak thermal disk component. We have shown recently \citep{Bursa04} that significant temporal variations in the observed flux can be accomplished by oscillations in the geometrically thick flows, fluid tori, even if they are axially symmetric. Here we propose that the QPO variations in the energetic part of the spectrum may come from such very hot and optically thin torus terminating the accretion flow, which exhibits two basic oscillating modes.

Relativistic tori will generally oscillate in a mixture of internal and global modes. Internal modes cause oscillations of the pressure and density profiles within the torus. The outgoing flux is therefore directly modulated by changes in the thermodynamical properties of the gas, while the shape of the torus is nearly unchanged, which is off our interest here. Global modes, on the other hand, alter mainly the spatial distribution of the material. Because light rays do not follow straight lines in a curved spacetime, these changes can be displayed out by effects of gravitational lensing and light bending.

In this paper we summarize extended results of numerical calculations and show how simple global oscillation modes of a gaseous torus affect the outgoing flux received by an static distant observer in the asymptotically flat spacetime and how the flux modulation depends on the geometry and various parameters of the torus. In Section~2 we briefly summarise the idea of the slender torus model and the equations, which are used to construct the torus and to set its radiative properties. In Section~3 we let the torus to execute global oscillations and using a numerical ray-tracing we inspect how these oscillations modulate the observed flux. If not stated otherwise, we use geometrical units \mbox{$c=G=1$} throughout this paper.

\section{Slender torus model}

The idea of a slender torus was initially invented by \citet{MadejPaczynski77} in their model of accretion disk of U~Geminorum. They noticed that in the slender limit (\ie when the torus is small as compared with its distance) and in the Newtonian potential, the equipotential surfaces are concentric circles. This additional symmetry induced by a Newtonian potential allowed \citet{Blaes85} to find a complete set of normal mode solutions for the linear perturbations of polytropic tori with constant specific angular momentum. He extended calculations done for a `thin isothermal ring' by \citet{PapaloizouPringle84} and showed how to find eigenfunctions and eigenfrequencies of all internal modes.

\citet{ABHKR05} have recently considered global modes of a slender torus and showed that between possible solutions of the relativistic Papaloizou-Pringle equation there exist also rigid and axisymmetric ($m\=0$) modes. These modes represent the simplest global and always-present oscillations in an accretion flow, axisymmetric up-down and in-out motion at the meridional and radial epicyclic frequencies.

\subsubsection*{Metric}

Most, if not all, of stellar and super-massive black holes have considerable amount of angular momentum, so that the Kerr metric has to be used to accurately describe their exterior spacetime. However, here we intend to study the basic effects of general relativity on the appearance of a moving axisymmetric body. We are mainly interested in how the light bending and gravitational lensing can modulate observed flux from sources. For this purpose we are pressing for a maximum simplicity to be able to isolate and recognise the essential effects of strong gravity on light.

Therefore, instead of the appropriate Kerr metric, we make use of the static Schwarzschild metric for calculations and where we compare with the non-relativistic case, the Minkowski flat spacetime metric is also used.

\subsubsection*{Equipotential structure}

The equipotential structure of a real torus is given by the Euler equation,
\begin{equation}
  \label{eq:euler}
  a_\mu = - \frac{\D{\mu} p}{p+\epsilon} \;,
\end{equation}
where \mbox{$a_\mu \!\equiv\! u^\nu\D{\nu} u_\mu$} is the 4-acceleration of the fluid and $\epsilon$, $p$ are respectively the proper energy density and the isotropic pressure. The fluid rotates in the azimuthal direction with the angular velocity $\Omega$ and has the 4-velocity of the form
\begin{equation}
  \label{eq:4-velocity}
  u^\mu = \big(u^t,\,0,\,0,\,u^\phi\big) = u^t\,\big(1,\,0,\,0,\,\Omega)
\end{equation}
After the substitution of \citeq{eq:4-velocity}, the Euler equation reads
\begin{equation}
  \label{eq:euler-2}
  - \frac{\D{\mu} p}{p+\epsilon} = \D{\mu}\,\Ueff - \frac{\Omega\D{\mu}\ell}{1-\Omega\,\ell} \;,
\end{equation}
where \mbox{$\Ueff\=-\frac12\ln\left(g^{tt} + \ell^2 g^{\phi\phi}\right)$} is the effective potential and $\ell$ is the specific angular momentum.

For a barotropic fluid, \ie the fluid described by a one-parametric equation of state $p\=p(\epsilon)$, the surfaces of constant pressure and constant total energy density coincide and it is possible to find a potential $W$ such that $W\=-\int_0^p \nabla{p}/(p+\epsilon)$, which simplifies the problem enormously \citep{AJS78}. 
The shape of the `equipotential' surfaces \mbox{$W(r,\,z)\=\text{const}$} is then given by specification of the rotation law \mbox{$\ell=\ell(\Omega)$} and of the gravitational field.

We assume the fluid to have uniform specific angular momentum,
\begin{equation}
  \label{eq:ell}
  \ell(r) = \ellK(r_0) = \frac{\sqrt{M\,r_0^3}}{r_0 - 2M} \;,
\end{equation}
where $r_0$ represents the centre of the torus. At this point, gravitational and centrifugal forces are just balanced and the fluid moves freely with the rotational velocity and the specific angular momentum having their Keplerian values $\OmegaK(r_0)$ and $\ellK(r_0)$. 

The shape of the torus is given by the solution of equation~\citeq{eq:euler-2}, which in the case of constant $\ell$ has a simple form, 
\begin{equation}
  W = \Ueff + \text{const} \;.
\end{equation}
In the slender approximation, the solution can be expressed in terms of second derivatives of the effective potential and it turns out that the torus has an elliptical cross-section with semi-axes in the ratio of epicyclic frequencies (\citealt{ABHKR05}; see also [\v{S}r\'{a}mkov\'{a}] in this proceedings).

\begin{figure}[t]
\resizebox{\hsize}{!}
{\includegraphics[height=5cm]{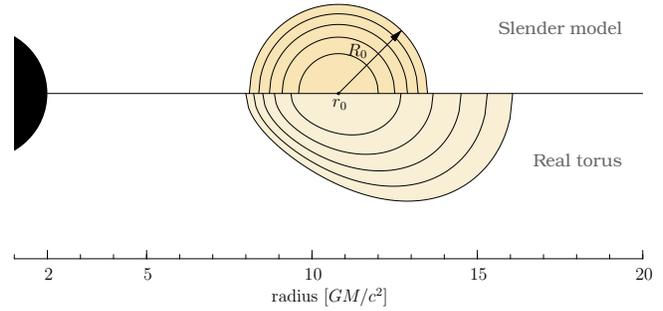}}
\caption{
An illustration of the equipotential structure of a real relativistic torus ({\em lower part}) and of our circular slender torus model ({\em upper part}) surrounding a black hole. The equipotential contours are separated by equal steps in the potential $W$.}
\label{fig:torus-equipotentials}
\end{figure}

In the model used here, we make even greater simplification. Introducing the cylindrical coordinates \mbox{$(t,\,r,\,z,\,\phi)$}, we use only the expansion at \mbox{$r\!=\!r_0$} in the \mbox{$z$-direction} to obtain a slender torus with a circular cross-section of equipotential surfaces,
\begin{equation}
  W(r,\,z) = \frac12 \ln\!\left[ \frac{(r_0-2M)^2}{r_0\,(r_0-3M)} \right] + \frac{M\,[(r\!-\!r_0)^2\!+\!z^2]}{2\,r_0^2\,(r_0-3M)} \,.\;\;
\end{equation}
The profiles of the equipotential structure of a relativistic torus and of our model are illustrated in Fig.~\ref{fig:torus-equipotentials}.

\subsubsection*{Thermodynamics}

An equation of state of polytropic type, 
\begin{equation}
  p=K\,\rho^\gamma \;,
\end{equation}
is assumed to complete the thermodynamical description of the fluid. Here, $\gamma$ is the adiabatic index, which have a value of $\ratio53$ for an adiabatic mono-atomic gas, and $K$ is the polytropic constant determining the specific adiabatic process.

\begin{figure}[b]
\resizebox{\hsize}{!}
{\includegraphics[height=5cm]{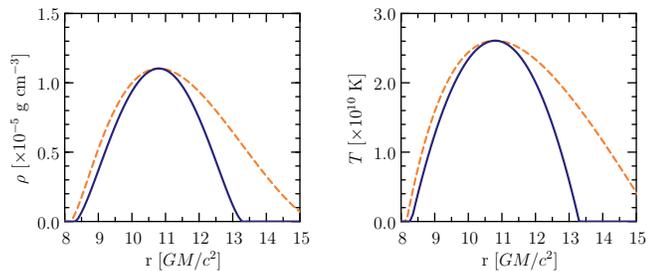}}
\caption{
The density ({\it left}) and temperature ({\it right}) profiles of a polytropic gas forming an accretion torus with the centre at \mbox{$r_0\!=\!10.8\,M$}. Solid lines represent the slender model with radius \mbox{$R_0\!=\!2\,M$} and dashed lines represent the real torus filling the potential well of the same depth.}
\label{fig:torus-rho-T}
\end{figure}

Now, we can directly integrate the right-hand side of the Euler equation \citeq{eq:euler} and obtain an expression for the potential W in terms of fluid density,
\begin{equation}
	W = \ln \rho - \ln\left(K\,\gamma\,\rho^\gamma + \rho\,\gamma - \rho \right) + \ln\left(\gamma-1\right) \;,
\end{equation}
where we have fixed the integration constant by the requirement \mbox{$W(\rho\=0)=0$}. The density and temperature profiles are therefore
\begin{align}
	\rho &= \left[ \frac{\gamma-1}{K\,\gamma} \left(e^W-1\right) \right]^\frac{1}{\gamma-1} \;,\\[0.5em]
	T &= \frac{\amu\,\mw}{\kB}\,\frac{p}{\rho} = \frac{\amu\,\mw}{\kB} \frac{\gamma-1}{\gamma} \left(e^W-1\right) \;,
\end{align}
where $\mw$, $\kB$ and $\amu$ and the molecular weight, the Boltzmann constant and the atomic mass unit, respectively (Fig.~\ref{fig:torus-rho-T}). 

\subsubsection*{Bremsstrahlung cooling \footnote{CGS units are used in this paragraph}}

We assume the torus to be filled with an optically thin gas radiating by the bremsstrahlung cooling. The emission include radiation from both electron\-ion and electron\-electron collisions \citep{StepneyGuilbert83, NarayanYi95}:
\begin{equation}
  f = f_{ei} + f_{ee} \;.
\end{equation}
The contributions of either types are given by
\begin{align}
  f_{ei}& = n_e\,\bar{n}\,\sigmathomson\,c\,\finestruct\,m_e\,c^2\,F_{ei}(\theta_{e}) \quad \text{and} \\
  f_{ee}& = n_e^2 c\,r_e^2 \finestruct\,m_e\,c^2 F_{ee}(\theta_{e}) \;,
\end{align}
where $n_e$ and $\bar{n}$ are number densities of electrons and ions, $\sigmathomson$ is Thomson cross-section, $m_e$ and \mbox{$r_e\!=\!e^2/m_e c^2$} denote mass of electron and its classical radius, $\finestruct$ is the fine structure constant, $F_{ee}(\theta_{e})$ and $F_{ei}(\theta_{e})$ are radiation rate functions and \mbox{$\theta_e\!=\!k\,T_e/m_e\,c^2$} is the dimensionless electron temperature. $F_{ee}(\theta_{e})$ and $F_{ei}(\theta_{e})$ are about of the same order, so that the ratio of electron\-ion and electron\-electron bremsstrahlung is
\begin{align}
  \frac{f_{ei}}{f_{ee}} \approx \frac{\sigmathomson}{r_e^2} \approx 8.4
\end{align}
and we can neglect the contribution from electron\-electron collisions. For the function $F_{ei}(\theta_{e})$ \citet{NarayanYi95} give the following expression:
\begin{align}
  F_{ei}(\theta_{e}) &= 4\left(\frac{2\theta_e}{\pi^3}\right)^{1/2} \left[1+1.781\,\theta_e^{1.34}\right] \;, 
  \quad &\theta_e<1 \;, \\
                     &= \frac{9\theta_e}{2\pi} \left[\ln(1.123\,\theta_e + 0.48) + 1.5\right] \;,
  \quad &\theta_e>1 \;.
\end{align}

In case of a multi-component plasma, the density $\bar{n}$ is calculated as a sum over individual ion species, \mbox{$\bar{n}\!=\!\sum Z_j^2\,n_j$}, where $Z_j$ is the charge of $j$-th species and $n_j$ is its number density. For a hydrogen\-helium composition with abundances $X\!:\!Y$ holds the following:
\begin{alignat}{3}
  n_e & \equiv \sum Z_j\,n_j & &=& 
      {\textstyle \frac{X+2\,Y}{X+Y}\,\sum n_j} \;, \\
  \bar{n} & \equiv \sum Z_j^2\,n_j & &=& 
      {\textstyle \frac{X+4\,Y}{X+Y}\,\sum n_j} \;, \\ 
  \rho & \equiv \sum {A_\text{r}}_j\,\amu\,n_j & &=&\;
      {\textstyle \amu\,\frac{X+4\,Y}{X+Y}\,\sum n_j} \;,
\end{alignat}
where ${A_\text{r}}_j$ is the relative atomic weight of the \mbox{$j$-th} species, $\amu$ denotes the atomic mass unit and we define \mbox{$\mu \equiv (X+4Y)/(X+Y)$}. The emissivity is then 
\begin{equation}
  f_{ei} = 4.30 \times 10^{-25}\,\tfrac{\mu+2}{3\,\mu}\,\rho^2\,F_{ei}(\theta_{e})\ \erg\,\cm^{-3}\,\s^{-1}\;,
\end{equation}
which for the non-relativistic limit ($\theta_e\!\ll\!1$) and Population~I abundances ($X\!=\!0.7$ and $Y=0.28$) gives
\begin{equation}
  \label{eq:emissivity}
  f_{ei} = 3.93 \times 10^{20}\,\rho^2\,T^\ratio12\ \erg\,\cm^{-3}\,\s^{-1}\;.
\end{equation}

\section{Torus oscillations}

\begin{figure*}[t!]
\resizebox{\hsize}{!}{
\includegraphics{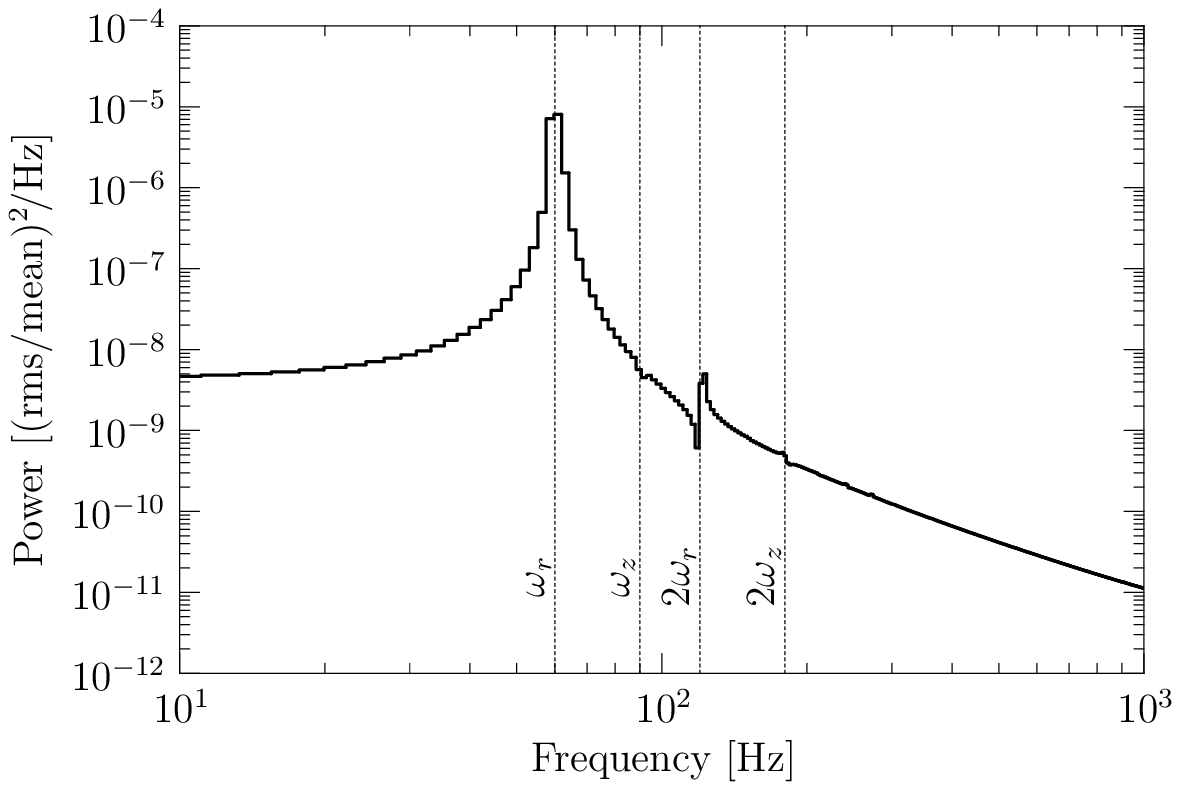}
\includegraphics{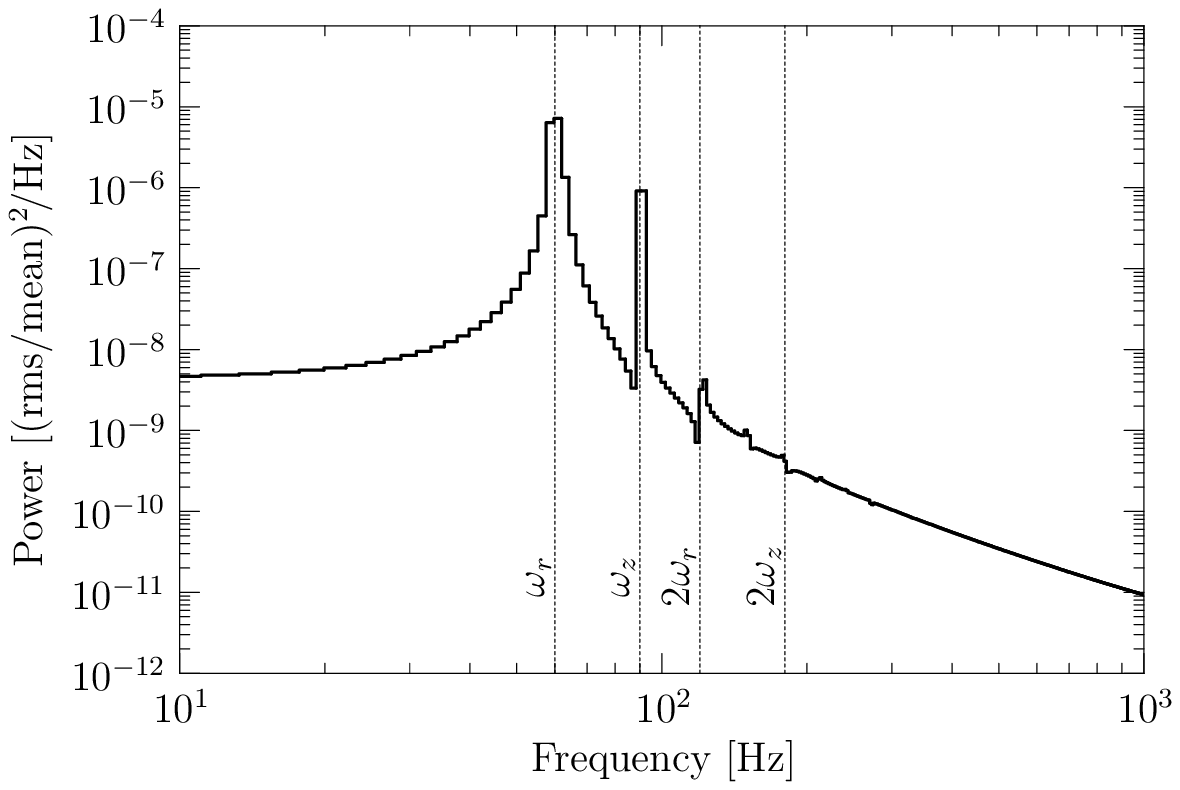}
\includegraphics{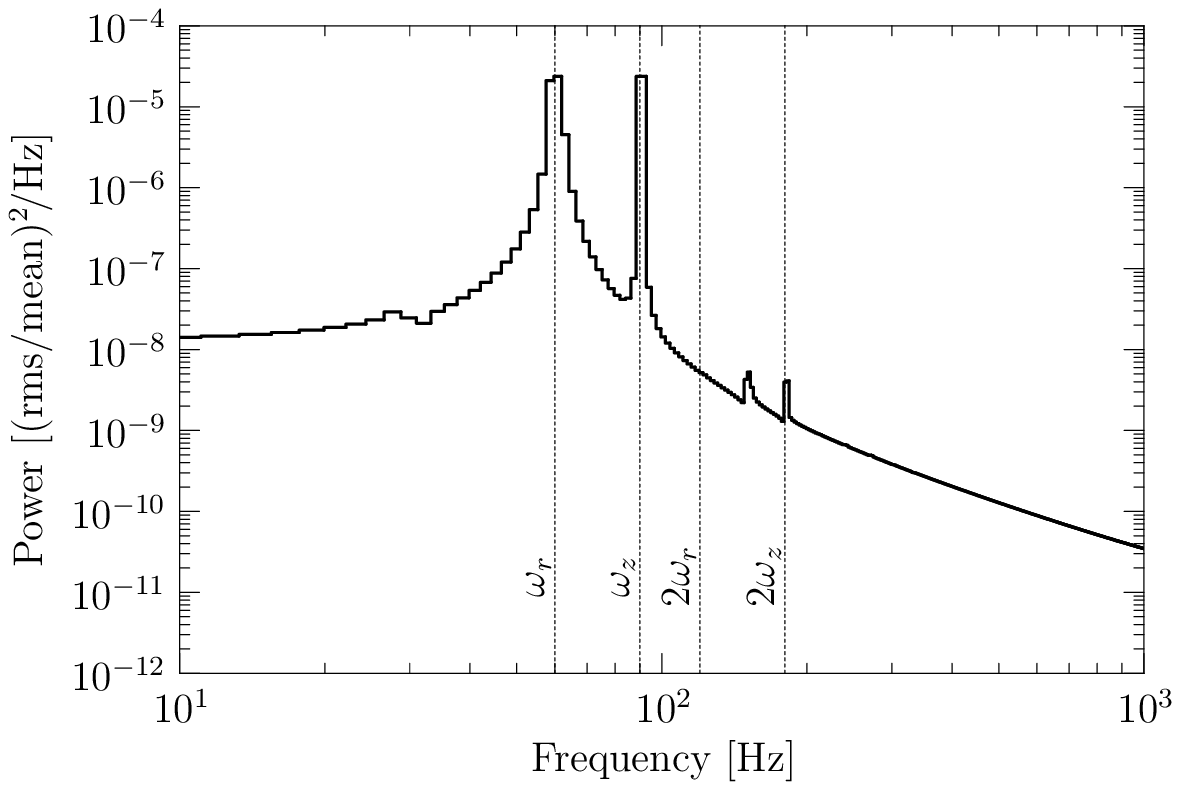}}
\caption{Power spectra of an oscillating torus calculated in the Newtonian limit ({\it left}), Minkowski spacetime ({\it middle}) and the Schwarzschild spacetime ({\it right}). Viewing angle is $70\deg$.}
\label{fig:effect-geometry}
\end{figure*}

\begin{figure}[t]
\resizebox{\hsize}{!}
{\includegraphics{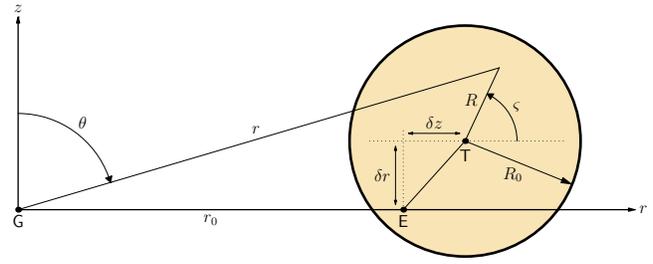}}
\caption{A schematic illustration of the displacement. The centre \textsf{T} of the torus is displaced radially by $\delta r$ and vertically by $\delta z$ from its equilibrium position \textsf{E}, which is at the distance $r_0$ from the centre of gravity \textsf{G}.}
\label{fig:torus-configuration}
\end{figure}

In the following, we excite in the torus rigid and axisymmetric \mbox{($m\!=\!0$)} sinusoidal oscillations in the vertical direction, \ie parallel to its axis, as well as in the perpendicular radial direction. Such an assumption will serve us to model the possible basic global modes found by \citet{ABHKR05}. In our model, the torus is rigidly displaced from its equilibrium (Fig.~\ref{fig:torus-configuration}), so that the position of the central circle varies as
\begin{equation}
r(t) = r_0 + \delta{r}\,\sin(\omega_r t) \;, \quad
z(t) = \delta{z}\,\sin(\omega_z t) \;.
\end{equation}
Here, \mbox{$\omega_z = \OmegaK=(M/r_0^3)^\frac12$} is the vertical epicyclic frequency, in Schwarzschild geometry equal to the Keplerian orbital frequency, and \mbox{$\omega_r = \OmegaK(1-6M/r_0)^\frac12$} is the radial epicyclic frequency. The torus is placed at the distance \mbox{$r_0\=10.8\,M$} so that the oscillation frequency ratio \mbox{$\omega_z:\omega_r$} is \mbox{$3:2$}, but the choice is arbitrary. If not stated otherwise, the cross-section radius is \mbox{$R_0\=2.0\,M$} and amplitudes of the both vertical and radial motion are set to \mbox{$\delta{z}=\delta{r}=0.1\,R_0$}.

We initially assume the `incompressible' mode, where the equipotential structure and the thermodynamical quantities describing the torus are fixed and do not vary in time as the torus moves. Later in this Section we describe also the `compressible' mode and discuss how changes in the torus properties affect powers in the different oscillations.

The radial motion results in a periodic change of volume of the torus. Because the optically thin torus is assumed to be filled with a polytropic gas radiating by bremsstrahlung cooling and we fix the density and temperature profiles, there is a corresponding change of luminosity \mbox{$L\!\propto\!\int\!f\,\d{V}$}, with a clear periodicity at $2\pi/\omega_r$. On the contrary, the vertical motion does not change the properties of the torus or its overall luminosity. We find that in spite of this, and although the torus is perfectly axisymmetric, the flux observed at infinity clearly varies at the oscillation frequency $\omega_z$. This is caused by relativistic effects at the source (lensing, beaming and time delay), and no other cause need to be invoked to explain in principle the highest-frequency modulation of X-rays in luminous black-hole binary sources.

\subsubsection*{Effect of spacetime geometry}

In the Newtonian limit and when the speed of light \mbox{$c\!\rightarrow\!\infty$}, the only observable periodicity is the radial oscillation. There is no sign of the $\omega_z$ frequency in the power spectrum, although the torus is moving vertically. This is clear and easy to understand, because the \mbox{$c\!\rightarrow\!\infty$} limit suppresses the time delay effects and causes photons from all parts of the torus to reach an observer at the same instant of time, so it is really seen as rigidly moving up and down giving no reason for modulation at the vertical frequency.

When the condition of the infinite light speed is relaxed, the torus is no longer seen as a rigid body. The delay between photons, which originate at the opposite sides of the torus at the same coordinate time, is \mbox{$\Delta{t} \simeq 2\,r_0/c\, \sin{\inc}$}, where $\inc$ is the viewing angle (\ie inclination of the observer). It is maximal for an edge-on view (\mbox{$\inc\=\ratio{\pi}{2}$}) and compared to the Keplerian orbital period it is \mbox{$\Delta{t}/\TK \simeq (2\pi^2\,r_0/r_g)^{-1/2}$}. It makes about 10\% at \mbox{$r_0\=10.8M$}. The torus is seen from distance as an elastic ring, which modulates its brightness also at the vertical oscillation frequency $\omega_z$ due to the time delay effect and the seeming volume change.

Curved spacetime adds the effect of light bending. Photons are focused by the central mass's gravity, which leads to a magnification of any vertical movement. Black hole is not a perfect lens, so that the parallel rays do not cross in a single point, but rather form a narrow focal furrow behind it. When the torus trench the furrow (at high viewing angles), its oscillations are greatly amplified by the lensing effect. This is especially significant in the case of the vertical oscillation, as the bright centre of the torus periodically passes through the focal line.

Figure~\ref{fig:effect-geometry} illustrates the geometry effect on three Fourier power density spectra of an oscillating torus. The spectra are calculated for the same parameters and only the metric is changed. The appearance of the vertical oscillation peak in the `finite light speed' case and its power amplification in the relativistic case are clearly visible.

\subsubsection*{Effect of inclination}

\begin{figure}[t]
\resizebox{\hsize}{!}{
\includegraphics{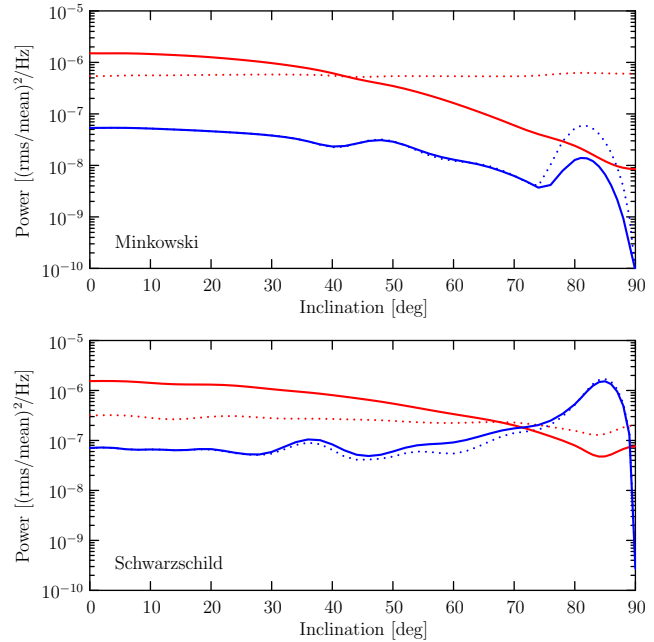}}
\caption{The inclination dependence of powers in the radial ({\it red}) and the vertical ({\it blue}) oscillations. Top panel shows calculations in the flat spacetime, bottom panel shows powers as computed in the curved Schwarzschild spacetime. Dashed lines represent the same calculations done with switched-off \mbox{$g$-factor} \mbox{($g \equiv 1$)}.}
\label{fig:effect-inclination-km0}
\end{figure}

\begin{figure}[t]
\resizebox{\hsize}{!}{
\includegraphics{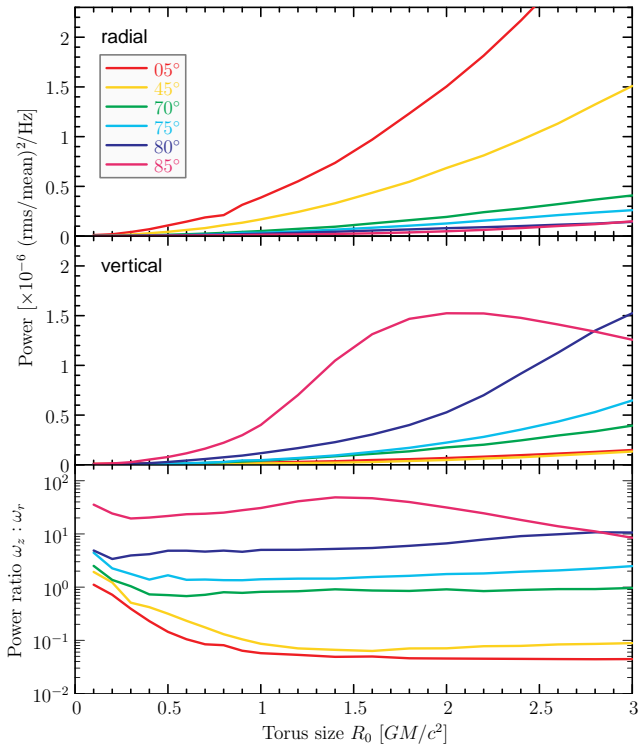}}
\caption{Powers in the radial ({\it top}) and vertical ({\it middle}) oscillations and their ratio ({\it bottom}) as a function of the torus size. Different viewing angles are plotted.}
\label{fig:effect-size-km0}
\end{figure}

\begin{figure}[t]
\resizebox{\hsize}{!}{
\includegraphics{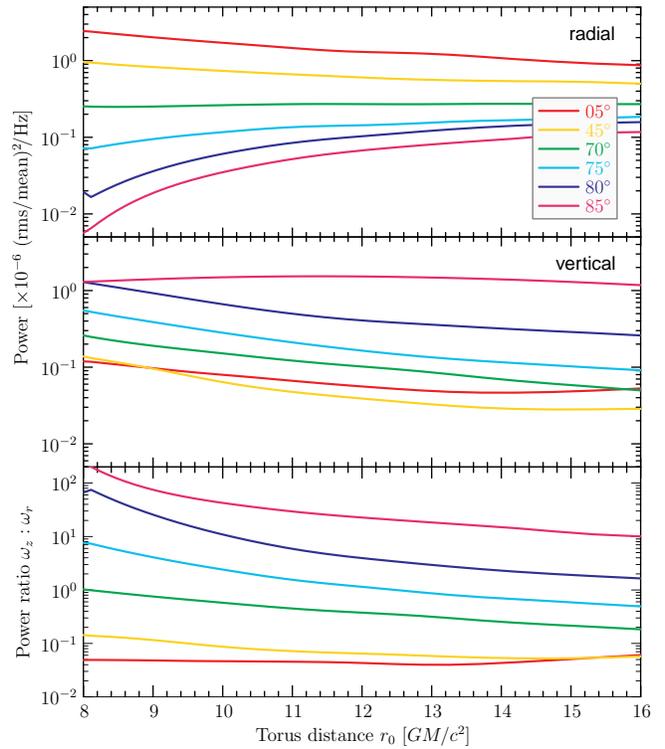}}
\caption{Powers in the radial ({\it top}) and vertical ({\it middle}) oscillations and their ratio ({\it bottom}) as a function of the torus distance from the gravity centre. Different viewing angles are plotted.}
\label{fig:effect-distance-km0}
\end{figure}

In previous paragraphs we have find out that both the time delay and the lensing effects are most pronounced when the viewing angle is rather high. Now we will show how much is the observed flux modulated when the torus is seen from different directions.

The effect of inclination is probably the most featured, in spite of it is difficult to be directly observed. Changing the line of sight mixes powers in amplitudes, because different effects are important at different angles. When the torus is viewed \mbox{face-on} (\ie from the top), we expect the amplitude of $\omega_r$ to be dominant, as the radial pulsations of the torus can be nicely seen and light rays passing through the gas are not yet strongly bended. When viewed almost edge-on, the Doppler effect dumps the power of $\omega_r$ and gravitational lensing amplifies the power in  $\omega_z$. Thus we expect the vertical oscillation to overpower the radial one.

Figure~\ref{fig:effect-inclination-km0} shows the inclination dependence of oscillation powers in the flat Minkowski spacetime ({\it top}) and in the curved Schwarzschild spacetime ({\it bottom}). We see that in the flat spacetime the power of radial oscillation gradually decreases, which is caused by the Doppler effect (\cf the red dotted line in the graph). The vertical oscillation decreases as well, but it is independent on \mbox{the $g$-factor}. At inclinations $\inc>75\deg$ it has a significant excess caused by the obscuration of part of the torus behind an opaque sphere of radius $2M$ representing the central black hole.

When gravity is added, the situation at low inclinations (up to \mbox{$\inc\!\simeq\!25\deg$}) is very much similar to the Minkowski case. The power of gravitational lensing is clearly visible from the blue line, \ie the vertical oscillation, progression. It is raising slowly for inclinations \mbox{$\inc\!>\!45\deg$}, then it shows a steeper increase for \mbox{$\inc\!>\!75\deg$}, reaches its maximum at \mbox{$\inc\!=\!85\deg$} and it finally drops down to zero. At the maximum it overpowers the radial oscillation by a factor of 40, while it is $20\times$ weaker if the torus is viewed \mbox{face-on}. The rapid decrease at the end is caused by the equatorial plane symmetry. If the line of sight is in the \mbox{$\theta\!=\!\ratio{\pi}{2}$} plane, the situation is the same above and below the plane, thus the periodicity is $2\,\omega_z$. The power in the base frequency drops abruptly and moves to overtones.

\subsubsection*{Effect of the torus size}

The effect of the size of the torus is very important to study, because it can be directly tested against observational data. Other free model parameters tend to be fixed for a given source (\eg like inclination), but the torus size may well vary for a single source as a response to temporal changes in the accretion rate.

The power in the radial oscillation is correlated with its amplitude, which is set to \mbox{$\delta{r}\!=\!0.1\,R_0$} and grows with the torus size. It is therefore evident, that the radial power will be proportional to $R_0$ squared. If the amplitude was constant or at least independent of $R_0$, the $\omega_r$ power would be independent of $R_0$ too. Thus the non-trivial part of the torus size dependence will be incurred by vertical movements of the torus.

Figure~\ref{fig:effect-size-km0} shows the PSD power profiles of both the radial and vertical oscillations for several different inclinations. Indeed, the radial power has a quadratic profile and is more dominant for lower viewing angles, which follows from the previous paragraph. The power in the vertical oscillation is at low inclinations also quadratic and similar to the radial one, but the reason is different. The time delay effect causes apparent deformations from the circular cross-section as the torus moves up and down, \ie to and from the observer in the case of a face-on view. The torus is squeezed along the line of sight at the turning points and stretched when passing the equatorial plane. Deformations are proportional to its size, being the reason for the observed profile. At high inclinations the appearance of strong relativistic images boosts the vertical oscillation power even more. But, as can be clearly seen from the $85\deg$ line and partially also from the $80\deg$ line, there is a size threshold, beyond which the oscillation power decreases though the torus still grows. This corresponds to the state, where the torus is so big that the relativistic images are saturated. Further increase of the torus size only entails an increase of the total luminosity, while the variability amplitude remains about the same, hence leading to the fractional rms amplitude downturn.

\subsubsection*{Effect of the torus distance}

The distance of the torus also affects the intensity of modulations in observed lightcurves (Fig.~\ref{fig:effect-distance-km0}). The power in the radial oscillation is either increasing or decreasing, depending on the inclination. Looking face-on, the $g$-factor is dominated by the redshift component and the power in $\omega_r$ is increasing with the torus distance being less dumped. When the view is more inclined, the Doppler component starts to be important and the oscillation looses power with the torus distance. The critical inclination is about $70\deg$. 

The power of vertical oscillation generally decreases with the torus distance. It is made visible mainly by the time delay effect and because with the increasing distance of the torus the oscillation period also increases, the effect is loosing on importance. An exception is when the inclination is very high. The large portion of visible relativistic images causes the vertical power first to increase up to some radius, beyond which it then decays. Both small and large tori do not have much of visible secondary images, because they are either too compact or they are too far. The ideal distance is about $11\,M$ -- this is the radius, where the torus has the largest portion of higher-order images, corresponding to the maximum of the vertical power in Fig.~\ref{fig:effect-distance-km0}.

Generally, the relative power of the vertical oscillation is getting weaker as the torus is more and more far-away from the graviting centre. This is most significant for higher viewing angles, where the drop between $8M$ and $16M$ can be more than one order of magnitude. On the other hand, for low inclinations the effect is less dramatic and if viewed face-on the power ratio is nearly independent from the distance of the fluid ring.

\subsubsection*{Effect of radial luminosity variations}

\begin{figure}
\resizebox{\hsize}{!}{
\includegraphics{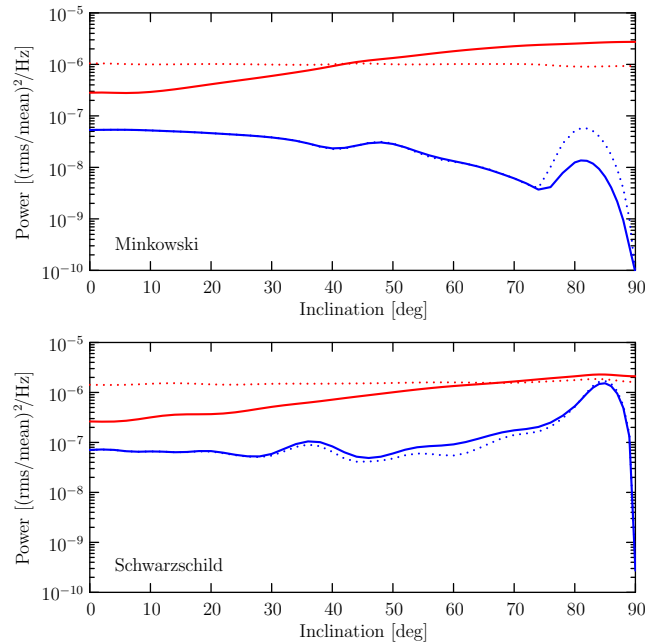}}
\caption{The inclination dependence of powers in the radial ({\it red}) and the vertical ({\it blue}) oscillations in the compressible mode. This is the same figure as Fig.~\ref{fig:effect-inclination-km1}, except that it is computed with the inclusion of density scaling. Top panel shows calculations in the flat spacetime, bottom panel shows powers as computed in the curved Schwarzschild spacetime. Dashed lines represent the same calculations done with switched-off \mbox{$g$-factor}.}
\label{fig:effect-inclination-km1}
\end{figure}

As already mentioned above, the volume of the torus changes periodically as the torus moves in and out. In the incompressible torus, which we have considered so far, this results in a corresponding variance of the luminosity, linearly proportional to the actual distance of the torus $r(t)$ from the centre,
\begin{equation}
  L(t) \sim {\textstyle \int f\,\d{V}} \sim r(t) \sim \delta{r}\,\sin(\omega_r t) \;.
\end{equation}
Because we do not change the thermodynamical properties, it also means that the total mass \mbox{$M\!=\!\int\!\rho\,\d{V}$} contained within the torus is not conserved during its radial movements, which is the major disadvantage. In this paragraph we relax this constraint and explore the compressible mass conserving mode.

A compressed torus heats up, which results in an increase of its luminosity and size. These two effects go hand-in-hand, however to keep things simple we isolate them and only show, how powers are affected if we only scale the density and temperature without changing the torus cross-section.

We allow the torus to change the pressure and density profiles in a way that it will keep its total mass constant. The volume element $\d{V}$ is proportional to $r$, so that in order to satisfy this condition, the density must be scaled as
\begin{equation}
  \rho(r,\,z,\,t) = \rho^\circ(r,\,z) \, \frac{r_0}{r(t)} \;,
\end{equation}
where $\rho^\circ$ refers to the density profile of a steady non-oscillating torus with central ring at radius $r_0$. If we substitute for the emissivity from \citeq{eq:emissivity}, we find out that the luminosity now goes with $r$ as
\begin{equation}
  L(t) \sim {\textstyle \int f(\rho)\,\d{V}} \sim {\textstyle \int \rho^{7/3}\,\d{V}} \sim r(t)^{-1.33} \;.
\end{equation}
The negative sign of the exponent causes the luminosity to increase when the torus moves in and compresses. Moreover, the luminosity variance is stronger than in the incompressible case, because of the greater absolute value of the exponent. 

Figure~\ref{fig:effect-inclination-km1} shows the inclination dependence of oscillation powers in the compressible case. Compared to Fig.~\ref{fig:effect-inclination-km0} we see that the signal modulation at vertical frequency is not affected, but the slope of the radial oscillation power is reversed. A key role in this reversing plays the $g$-factor, which combines effects of the Doppler boosting and the gravitation redshift. 

\begin{figure}
\resizebox{\hsize}{!}{
\includegraphics{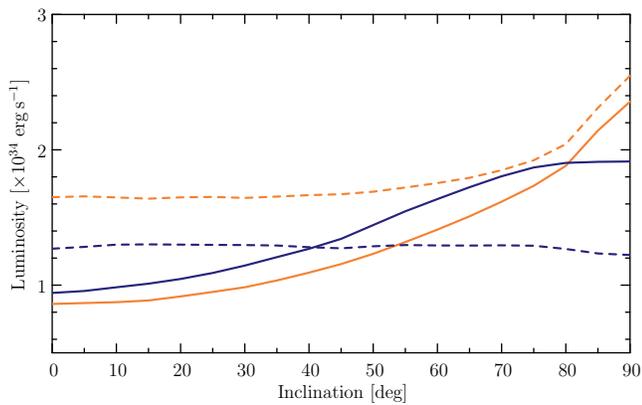}}
\caption{The total observed bolometric luminosity of a steady (non-oscillating) torus as a function of inclination. In a flat spacetime ({\it orange}) with only special relativistic effects, the total luminosity is increased by a factor of two if the view is changed from face-on to edge-on. It is even more in a curved spacetime ({\it blue}), where the relativistic images make a significant contribution. For comparison also calculations with switched-off \mbox{$g$-factor} (with $g$ being set to unity) are shown ({\it dashed} lines).}
\label{fig:lx-total-inclination}
\end{figure}

The Doppler effect brightens up the part of the torus, where the gas moves towards the observer, and darkens the receding part. This effect is maximal for inclinations approaching $\ratio{\pi}{2}$, \ie for \mbox{edge-on} view. On average, \ie integrated over the torus volume, the brighten part wins and the torus appears more luminous when viewed edge-on (see Fig.~\ref{fig:lx-total-inclination}).

The redshift effect adds the dependence on the radial distance from the centre of gravity, which is an important fact to explain the qualitative difference between Figs.~\ref{fig:effect-inclination-km0} and \ref{fig:effect-inclination-km1}. In the incompressible mode, the luminosity has a minimum when the torus moves in and a maximum when it moves out of its equilibrium position. The \mbox{$g$-factor} goes the same way and consequently amplifies the amplitude of the luminosity variability. The situation is right opposite in the compressible mode and the luminosity has a maximum when the torus moves in and a minimum when it moves out. The \mbox{$g$-factor} goes with the opposite phase and dumps the luminosity amplitude. Because the difference in the \mbox{$g$-factor} value is more pronounced with inclination, it results in increasing or decreasing dependence of the radial power on inclination in the compressible or incompressible case, respectively.

\section{Discussion and Conclusions}

We have found out that intrinsic variations of the radiation emitted from inner parts of an accretion flow may be significantly modified by effects of a strong gravitational field. Above all we have shown that orientation of the system with respect to the observer is an important factor, which may alter the distribution of powers in different modes. However this effect, although strong, cannot be directly observed, because the inclination of a given source is fixed and mostly uncertain.

Within the model there are other parameters, which may be used for predictions of powers in different frequencies. We have shown that the size of the torus affects the power of the vertical oscillation. In this model this corresponds to an emission of harder photons from a hotter torus and provides a link between the model and observations. From those we know \citep{Remillard02} that the higher HFQPO peak is usually more powerful than the lower one in harder spectral states, which is consistent with the model, but the exact correlation depends on amplitudes of both oscillations.

The power in the radial oscillation very much depends on the thermodynamical properties of the torus and on its behaviour under the influence of radial movements. We have shown that different parametrizations of intrinsic luminosity in the \mbox{in-and-out} motion  (\ie compressible and incompressible modes) change power of the radial oscillation. On the other hand, the power of the vertical oscillation remains unaffected. This is an important fact and it means that the flux modulation at the vertical frequency is independent on the torus properties, driven by relativistic effects only.

Another model parameter is the distance of the thin accretion disk. The Shakura-Sunyaev disk is optically thick and blocks propagation of photons, which cross the equatorial plane at radii beyond its moving inner edge. Most of the stopped photons are strongly lensed and carry information predominantly about the vertical mode, thus the presence or not-presence of an opaque disk may be important for the power distribution in QPO modes. However, this effect is beyond the scope of this article and will be described in a separate paper.

\acknowledgements
I am thankful to all my collaborators and especially to M.~Abramowicz, V.~Karas and W.~Klu{\' z}niak for incentive comments. This work was supported by the Czech GAAV grant IAA~300030510. The Astronomical Institute is operated under the project AV0Z10030501.

\end{document}